\documentclass[fleqn,usenatbib]{mnras}
\usepackage{newtxtext,newtxmath}
\usepackage{geometry}    
\usepackage{setspace} 
\usepackage[parfill]{parskip}    
\usepackage{graphicx}
\usepackage{amssymb}

\usepackage{epstopdf}
\usepackage{natbib}
\usepackage{geometry,natbib,float,color}
\usepackage{gensymb}
\def\etal{{\it et al}}
\def\deg{^{\circ}}
\def\P3hat{{\mathaccent 94 P}_3}

\def\eg{{\it e.g.}}
\def\ie{{\it i.e.}}

\def\aap{A\&A}
\def\aaps{A\&A Suppl.}
\def\apj{Ap.J.}
\def\apjs{Ap.J. Suppl.}
\def\mnras{M.N.R.A.S.}
\def\nat{Nature}
\def\apss{Ap\&SS}

\title[Mode-Switching Phenomenon in Pulsar B0329+54]{Investigation of the Mode-Switching Phenomenon in Pulsar B0329+54 through Polarimetric Analysis}

\author[Brinkman, Mitra \& Rankin]{Casey Brinkman,$^{1, 2}$
\thanks{E-mail: clbrinkm@hawaii.edu}
Dipanjan Mitra,$^{1,3,4}$ \&
Joanna Rankin,$^{1}$
\\
$^{1}$Physics Department, University of Vermont, Burlington, Vermont 05401, USA \\
$^{2}$Institute for Astronomy, University of Hawai'i at Manoa, Honolulu, Hawaii 96822, USA \\
$^{3}$National Centre for Radio Astrophysics, Ganeshkhind, Pune 411007, India \\
$^4$Janusz Gil Institute of Astronomy, University of Zielona G\'ora, ul. Szafrana 2, 65-516 Zielona G\'ora, Poland
}                                   \date{Accepted XXX. Received YYY; in original form ZZZ}      

\begin{document}
    \maketitle
\begin{abstract} 
The phenomenon of profile mode switching in pulsars, where the stable average pulse profile changes to another stable state on the timescale of a pulsar's period, remains poorly understood. We sought to understand how pulsars undergo profile mode switching through a comparative analysis of the polarization and geometry of the two different profile modes of PSR B0329+54. The polarization behavior and fitted parameters of the rotation-vector model remain constant between modes, and the emission height remains constant as well. These similarities lend support to a model of pair production in the surface plasma that would cause a change in the available electrons and therefore the differential emission intensity. \\
\end{abstract}


\newpage

\label{firstpage} 

\section{Introduction}
Pulsars---rapidly rotating and highly magnetized neutron stars---are unipolar inductors, and the regions around the neutron stars can sustain a force-free corotating magnetosphere populated with a dense electron-positron plasma. This pair plasma is thought to be created in a local gap or unscreened electric field region in the magnetosphere. The magnetosphere has closed and open dipolar field-line regions, and a pulsar's coherent radio emission is thought to be excited due to the growth of plasma instabilities in the relativistic electron-positron plasma flowing along the open dipolar magnetic field lines. The individual pulses from a pulsar may assume radically different forms, but when several hundreds to thousands of pulses are averaged they then produce a stable pulse profile. Most pulsars form the same average profile over any period of time; however, some pulsars exhibit more than one "mode" wherein its average profile assumes several distinct forms due to different populations of individual pulses. The phenomenon of profile mode switching was first observed in PSR B1237+25 by Backer (1970) and soon thereafter in PSR B0329+54 by Lyne (1971).
\begin{figure*}
\begin{center}
\includegraphics[width=1.0\textwidth]{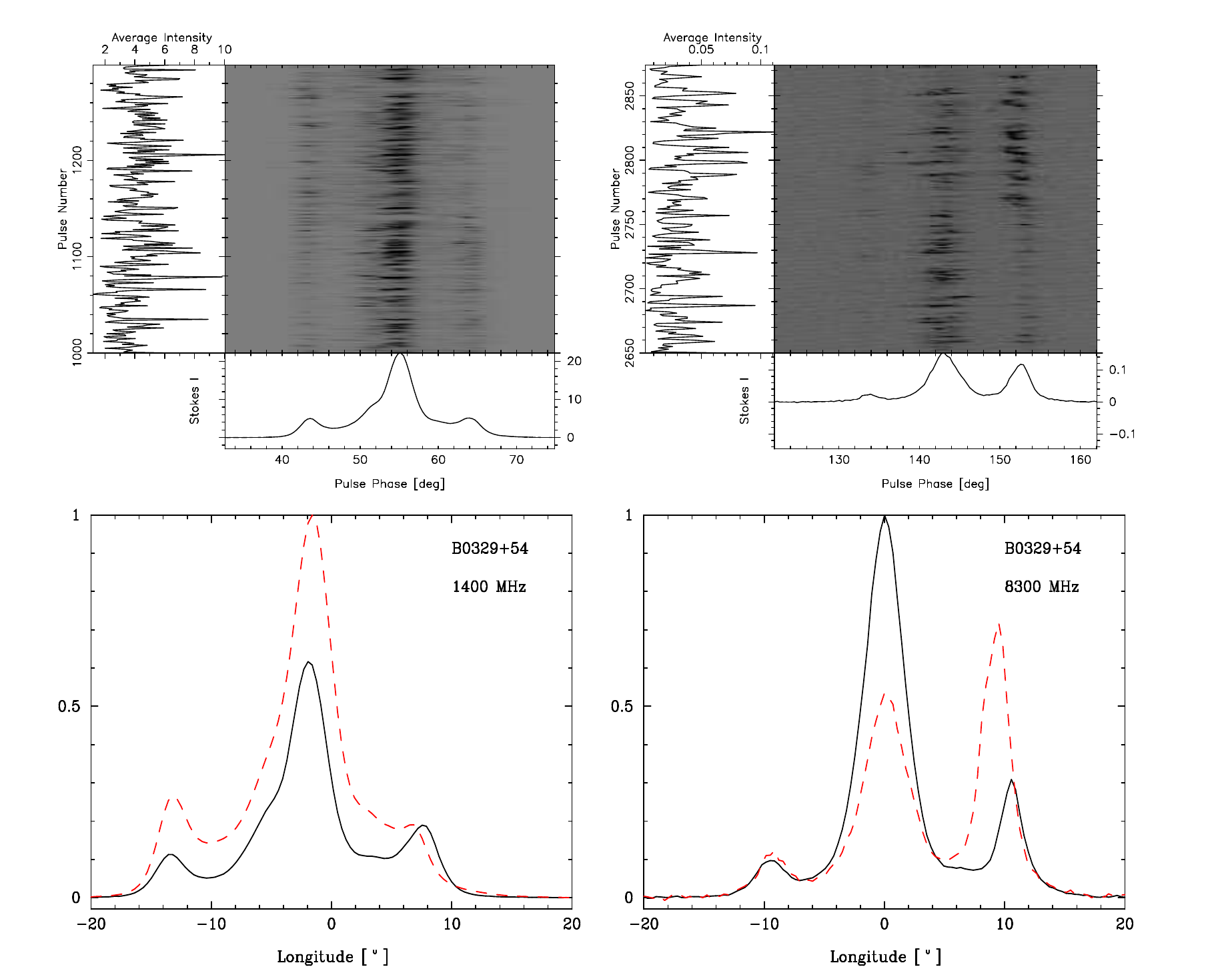} 
\end{center}
\caption{Normal {\it vs.} abnormal-mode multifrequency time series and
  profile shapes.  The top panels show the single pulse sequence at
  1400 and 8300 MHz where the mode switch can be seen at around pulse
  number 1120 and 2750 respectively.  The bottom panels show the
  average profile shape which appears quite different for the abnormal
  mode at different frequencies. The plot on the left shows the
  overlapping normal (solid black) and abnormal (dashed red) modes
  at 1400 MHz. The plot to the right shows the same at 8300 MHz. 
  The longitude zero for the 1400 MHz plot corresponds to the phase of the steepest
  gradient of the PPA obtained by fitting the RVM as discussed in section 3.2, 
  while for 8300 MHz the peak of the core of the normal mode is used.
  At 1400 MHz 
  the abnormal mode is about 70\% brighter than the normal mode; at 
  8300 MHz the intrinsic intensities of the two profiles cannot be reliably 
  determined because of refractive scintillation (see text).}
\label{fig1}
\end{figure*}

The physical mechanism responsible for mode switching, however, has remained mysterious. The phenomenon has been primarily associated with normal longer period pulsars (\ie, pulsars lying in the period range of a hundred milliseconds or longer), although mode changing has also been reported for some millisecond pulsars (\eg, Kramer
\etal\ 1999, Mahajan \etal\ 2018). The mode switching timescale can be
within a pulse period or over a few rotations, and the pulsar typically
remains in a particular mode for several minutes to hours. In some pulsars like PSRs B0031--07 (Smits \etal\ 2005) and J1822--2256
(Basu \& Mitra 2018), the pulsar can switch between several stable
modes, including a null state (where no radio emission is seen). Nulling timescales are similar to those of moding; however, an extreme
nulling timescale of weeks or months has been observed in intermittent
pulsars such as PSR B1931+24 (Kramer \etal\ 2006). 

Only recently have theorists attempted to understand moding effects in terms of a pulsar's global magnetospheric current configuration.  In the presence of copious pair-plasma creation, the global current flow in the pulsar magnetosphere can be modeled; however, time-dependent local pair creation processes can alter the global magnetospheric structure.  Timokhin (2010) suggests that these global magnetospheric
changes can manifest as different viewing geometries and/or current distributions in a given pulsar, both of which can produce the visible effects of mode changing.  Alternatively, Yuen \& Melrose (2017)
suggest that mode changing can reflect transitions of non-corotating to corotating states of a pulsar's magnetosphere. On the other hand, Cordes (2013) suggested that state change in pulsars is related to 
Markov processes in which state changes occur stochastically, and are decoupled from the global magnetospheric changes.
While a consensus on what causes mode changing in pulsars has yet to be reached, 
there are observational and analytical methods at our disposal to 
investigate the phenomenon.

Fortunately, we begin this work on a strong foundation.  Pulsar B0329+54 is one of the very brightest radio pulsars in the northern sky and an exemplar of the moding phenomenon.  The pulsar's average profile and polarization suggest that it has a five-component (M class) profile with a central core component and nested inner and outer conal emission.  Since moding was discovered in this pulsar, several detailed, long-term, and multifrequency studies have revealed that the mode changing is a broad band phenomenon that occurs simultaneously across frequencies, such that B0329+54 spends about 15\% of the time in its abnormal mode and the rest in its normal
mode (see Bartel \etal\ 1982, Yan \etal\ 2018, Chen \etal\ 2011, Bia\l{}kowski \etal\ 2018).

Most of these pulsar moding studies, except for Bartel \etal\ (1982), were carried out with total intensity.  A few detailed single pulse studies exist for this pulsar's normal mode (McKinnon \& Hankins 1993; Gil \& Lyne 1995; Edwards \& Stappers 2004; Mitra, Rankin \& Gupta 2007, hereafter MRG07).  Only Gil \& Lyne and MRG07 were polarimetric, where the burden was to understand the complex polarization properties of this pulsar, particularly the peculiar average linear polarization position-angle (PPA) traverse across the pulse.  Usually, the PPA exhibits an S-shaped traverse across the profile, which can be understood using the rotating vector model (RVM, see Radhakrishnan \& Cooke 1969), as evidence for the emission arising from open dipolar magnetic field lines centered around the magnetic dipole axis.  Gil \& Lyne (1995) and MRG07, based on a detailed single pulse analysis, showed that the complexity in the average PPA mostly occurred due to the averaging of the orthogonal polarization modes (OPM), which could largely be decoupled in the single pulse analysis as two orthogonal PPA tracks.  MRG07 also showed that if only single
pulses with weak or no core-emission were used, then the two orthogonal tracks were highly compatible with the RVM. This allowed the determination of two important features of the pulsar's emission: first, the aberration-retardation technique (\eg, Blaskiewicz \etal\ 1991; Mitra \& Li 2004) could be used to find the conal emission heights, and second, the dominant or the primary PPA track could be associated with the extraordinary (X) propagation mode and the weaker or secondary PPA track with the ordinary (O) propagation mode using the proper motion and absolute fiducial PPA measurements (see Johnston \etal\ 2005; Rankin 2007/2015; MRG07 for details).

MRG07 also found that when the single pulses with strong core emission were included, then a ``kink''-like feature appeared on top of the RVM-like PPA traverse.  The core emission has a intensity-dependent behaviour wherein the stronger core emission is seen ever earlier towards the leading part of the component. MRG07 attributed this effect to the higher intensity core emission arising from higher emission heights.  Thus, the A/R effect can both explain the intensity-dependent core emission and the PPA kink. These methods then allowed for the determination of the physical emission heights of the core and conal radiation of B0329+54.

Because pulsar B0329+54 has such illuminating and well investigated polarization behavior, it provides a particularly transparent context in which to compare the radio emission of the pulsar's normal and abnormal modes: the goal of our analysis below.  For this purpose we use archival B0329+54 observations from the Effelsberg radio telescope, where both the normal and abnormal modes were detected in polarimetric single pulse sequences. In section 2 we describe the set of observations and the analysis methods, and in section 3 we outline the detailed investigation carried out to compare the radio emission properties in the two modes.  We then summarize the results in section 4.

\begin{figure*}
\begin{center}
\includegraphics[width=1.0\textwidth]{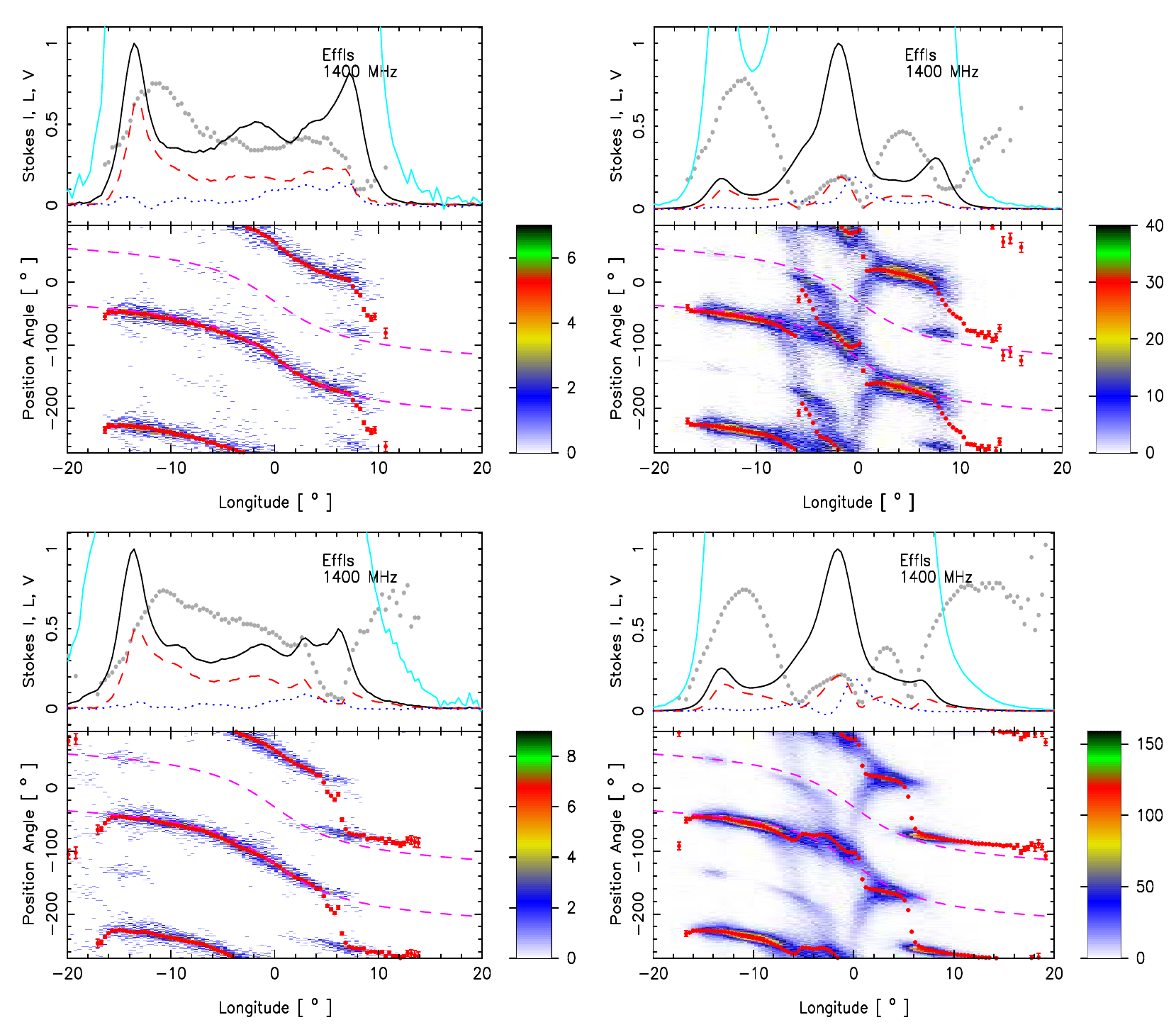}
\caption{Normal {\it vs} abnormal polarization at 1400 MHz.  Low
  (left) and full (right) intensity, average polarization
  displays at 1400 MHz for the normal (upper) and abnormal
  (lower) modes.  Each display gives the total
  intensity (Stokes $I$; solid curve), the average linear ($L$; dashed
  red), the circular polarization (Stokes $V$; dotted blue), the
  fractional linear polarization ($L/I$; gray points) and a zoomed
  total intensity (10$\times I $; cyan curve).  The PPA histogram
  (lower panels) within each 1$\degr$x1-sample cell correspond to
  samples having errors less than 3 $\sigma$ in $L$, are plotted
  according to the color bars at the lower right.  The average PPA
  traverses (red) are overplotted, and the RVM fit to the PPA traverse
  is plotted twice for the two polarization modes (magenta dashed) on
  each panel.  The origin is taken at the fitted PPA inflection
  point.}

\end{center}
\label{fig2}
\end{figure*}

\section{Observations} 
Abnormal mode intervals are infrequent in B0329+54's pulse sequences, and can be difficult to distinguish at lower frequencies. Therefore, for this work we obtained archival full Stokes single pulse observations made with the Effelsberg Radio Telescope at 8300, 5000, 2700, and 1400 MHz using the Observatory's earlier backend Effelsberg Pulsar Observing System (EPOS, Jessner 1996).  The data were available in the EPOS-format, and we used codes (provided to us by Ramesh Karuppusamy, private communication) to convert them to ascii formats. The full Stokes single pulse data had the pulsed calibration signals as described in von Hoensbroeck and \& Xilouris (1997), and using the calibration technique described in von Hoensbroeck (1999) we wrote our own software to obtain calibrated Stokes parameters $I,Q,U$ and $V$.
 
The pulse sequences were then further analyzed in a time-averaged as well as single pulse manner. The average total intensity profile across profile phase $\varphi_i$, was obtained as:
\begin{center}
$I(\varphi_i)=\sum_{j=1}^{n} I(\varphi_i)/n$ 
\end{center}
where $n$ is the total number of pulses. The average linear polarization $L(\varphi_i)$ was obtained by summing up Stokes $Q(\varphi_i)$ and $U(\varphi_i)$ along each $\varphi_i$, and using the relation: 
\begin{center}
$L(\varphi_i) =\sqrt{(\sum_{j=1}^{n}U_j(\varphi_i))^2 + (\sum_{j=1}^{n}Q_j(\varphi_i))^2}/n$   
\end{center}
The $L(\varphi_i)$ estimated above has a positive bias, and a mean value of the linear polarization obtained from the off pulse region is subtracted to obtain the final $L(\varphi_i)$. The average circular polarization was obtained using the relation: 
\begin{center}
$V (\varphi_i) = \sum_{j=1}^{n}V_j(\varphi_i)/n$ 
\end{center}
The average degrees of linear and circular polarization are defined as:
\begin{center}
\%$L  = \sum_{i} L(\varphi_i)/I(\varphi_i)$ 
\end{center}
and 
\begin{center}
\% $V = \sum_{i} V(\varphi_i)/I(\varphi_i)$ 
\end{center}
where the summation across pulse phase is performed for statistically significant values with $S/N > 3$, and the noise was estimated from the off-pulse region. The error in these average quantities was estimated using the methods
described in Mitra et al. (2016). The average polarization position angle was obtained as:
\begin{center}
$\chi(\varphi_i) = 0.5\tan^{-1} (\sum_{j=1}^{n}U_j(\varphi_i)/\sum_{j=1}^{n}Q_j(\varphi_i))$
\end{center}
with only points greater than three times the rms of the linear polarization baseline level being used.  The error in $\chi(\varphi_i)$ was calculated by propagation of errors of $U_j(\varphi_i)$ and $Q_j(\varphi_i)$ with their error obtained from their baseline level.

The width of the average profile was measured in various ways: namely, $W_{3\sigma}$ widths were measured using points on the profile corresponding to three times the rms values at the leading and trailing edges of the profile, the $W_{10}$ and $W_{50}$ correspond to the width measured by using 10\% and 50\% intensity points of the outer leading and trailing components of the profile, and the errors in the widths are computed using the method of Kijak \& Gil (1997). To obtain the linear, circular polarization and PPA for single pulses, average profiles were created by setting $n=1$.

\begin{figure*}
\begin{center}
\includegraphics[width=1.0\textwidth]{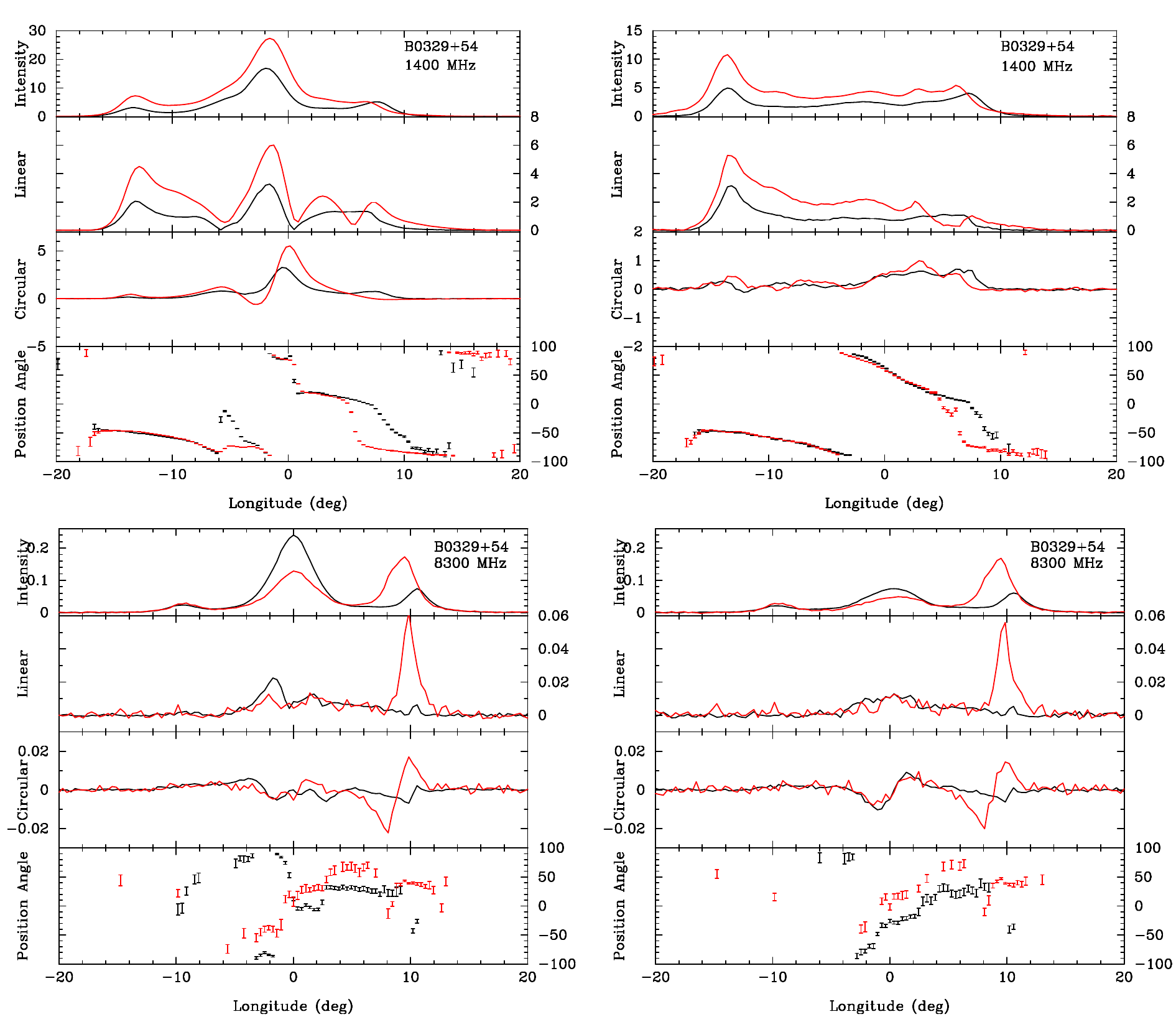}
\caption{Normal {\it vs} abnormal Stokes parameters at 1400 MHz (top) and 
8300 MHz (bottom).  The four
  images in each panel above provide a direct comparison of the intensity, linear
  polarization, circular polarization, and polarization position angle
  between the modes. The normal mode quantities are plotted in black
  while the abnormal ones are shown in red. The left panel averages the full 
pulse sequence, while the right panel shows the low core-intensity profiles (see
text for details). }
\end{center}
\label{fig3}
\end{figure*}

\begin{table}
\begin{tabular}{|l|l|l|}
\hline
PSR B0329+54    &         &          \\
\hline
Period (s)  & 0.714 & \\
$\dot{\rm{P}}$ &  $2.05 \times 10^{-15}$  \\ 
DM (pc/cc)  & 26.76 \\
RM rad m$^{-2}$ & -63.7 \\
             & &\\
Effelsberg Obs  &  Normal & Abnormal \\
Frequency (MHz) & 1400 &     \\
MJD            & 51971 &     \\
Number of Pulses & 1-1000 & 1000-3000\\
$W_{3\sigma}$ & 39.7$\pm$0.4\degr & 41.8$\pm$0.4\degr \\
$W_{10\%}$    & 27.1$\pm$0.2\degr & 27.6$\pm$0.2\degr \\
$W_{50\%}$    &23.9$\pm$0.2 \degr & 22.9$\pm$0.2\degr \\
{\%L}         &22.8$\pm$0.1       &26.7$\pm$0.1   \\
{\% V}        &14.6$\pm$0.1       &11.3$\pm$0.1   \\
             & &\\
Frequency (MHz) & 8300 &     \\
MJD            & 53184  &     \\
Number of Pulses & 1-2700 & 2700-3000\\
$W_{3\sigma}$ & 31.3$\pm$0.4\degr            &     28.2$\pm$0.4\degr           \\
$W_{10\%}$    & 26.4$\pm$0.2\degr            &     25.8$\pm$0.2\degr           \\
$W_{50\%}$    & 22.5$\pm$0.2\degr             &     21.1$\pm$0.2\degr     \\
{\%L}         & 9.8$\pm$0.5               &     19.2$\pm$2       \\
{\% V}        & -0.9$\pm$0.3               &     1.8$\pm$1.2       \\
             & &\\
\hline
\end{tabular}
\caption{Characteristics of the Effelsberg Observations}
\label{tab1}
\end{table}

\section{Analysis}
\subsection{Profile Mode Differentiation}

We identified intervals of mode changing in the 1400-MHz observations as well as those at 8300 MHz, and measured the relative amplitudes of the five components in order to quantify the differences between the normal and abnormal modes. The distinction between the normal and abnormal modes is most evident in the intensity of the trailing component relative to the other components: at lower frequencies (300 MHz-1400 MHz) the trailing component diminishes, whereas at higher frequencies (8300 MHz) it increases (see Bartel \etal\ 1981).

The top panels of Figure~\ref{fig1} show transitions from the normal to abnormal modes at both 1400 (left) and 8300 MHz (right). These mode switches occur within about a stellar rotation period (0.714 s), and the pulsar can remain stably in a particular mode for up to thousands of pulses.  By looking at the average profiles, we can see that at 1400 MHz the trailing component shrinks from 28\% the intensity of the main pulse in the normal mode to about 19\% in the abnormal.  At 8300 MHz, the trailing component grows from 33\% of the main pulse strength to 122\%.  The normal and abnormal mode average profiles are given in the lower panels of Figure~\ref{fig1}.

\begin{figure}
\begin{center}
\includegraphics[width=0.5\textwidth]{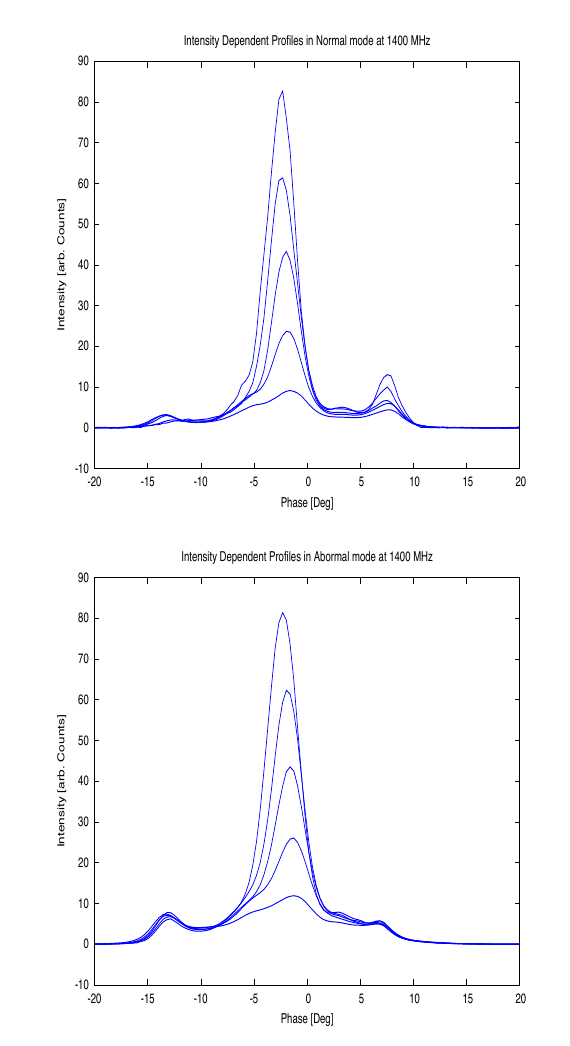}
\caption{Normal (upper) {\it vs} abnormal (lower) mode intensity-fractionated profiles at 1400 MHz. Individual pulses are separated
  into 5 different fractions based on their intensity and the averaged
  fractions plotted superimposed so that the angular shifts between
  the highest and lowest intensity fractions can be clearly seen and
  measured at some 2\degr.}
\end{center}
\label{fig4}
\end{figure}

\begin{figure*}
\begin{center}
\includegraphics[width=1.0\textwidth]{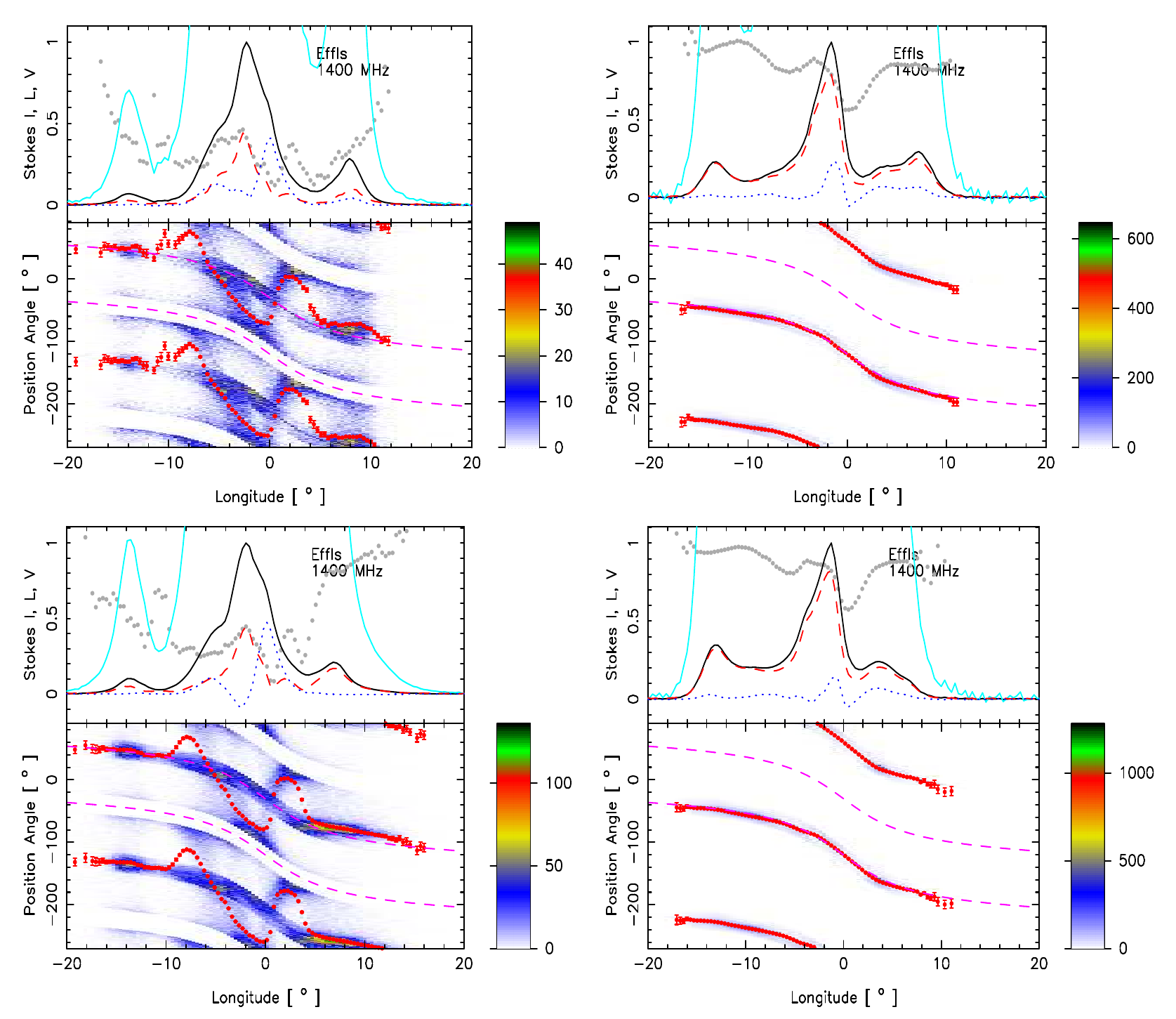}
\caption{Polarization mode-segregated profile displays for the 1400-MHz observation, normal
  mode (top row) and abnormal mode (bottom row), after Fig. 2 following the two-way technique described in Deshpande \& Rankin
  (2001).  The displays show the aggregate primary polarization mode (PPM, X, left), and secondary polarization mode (SPM, O, right).  The abnormal mode is 70\% stronger than the normal mode, and the X/O
  power ratio is 71\%/29\% in both modes.}

\end{center}
\label{fig5}
\end{figure*}

\begin{figure}[t]
\begin{center}
\includegraphics[width=0.5\textwidth]{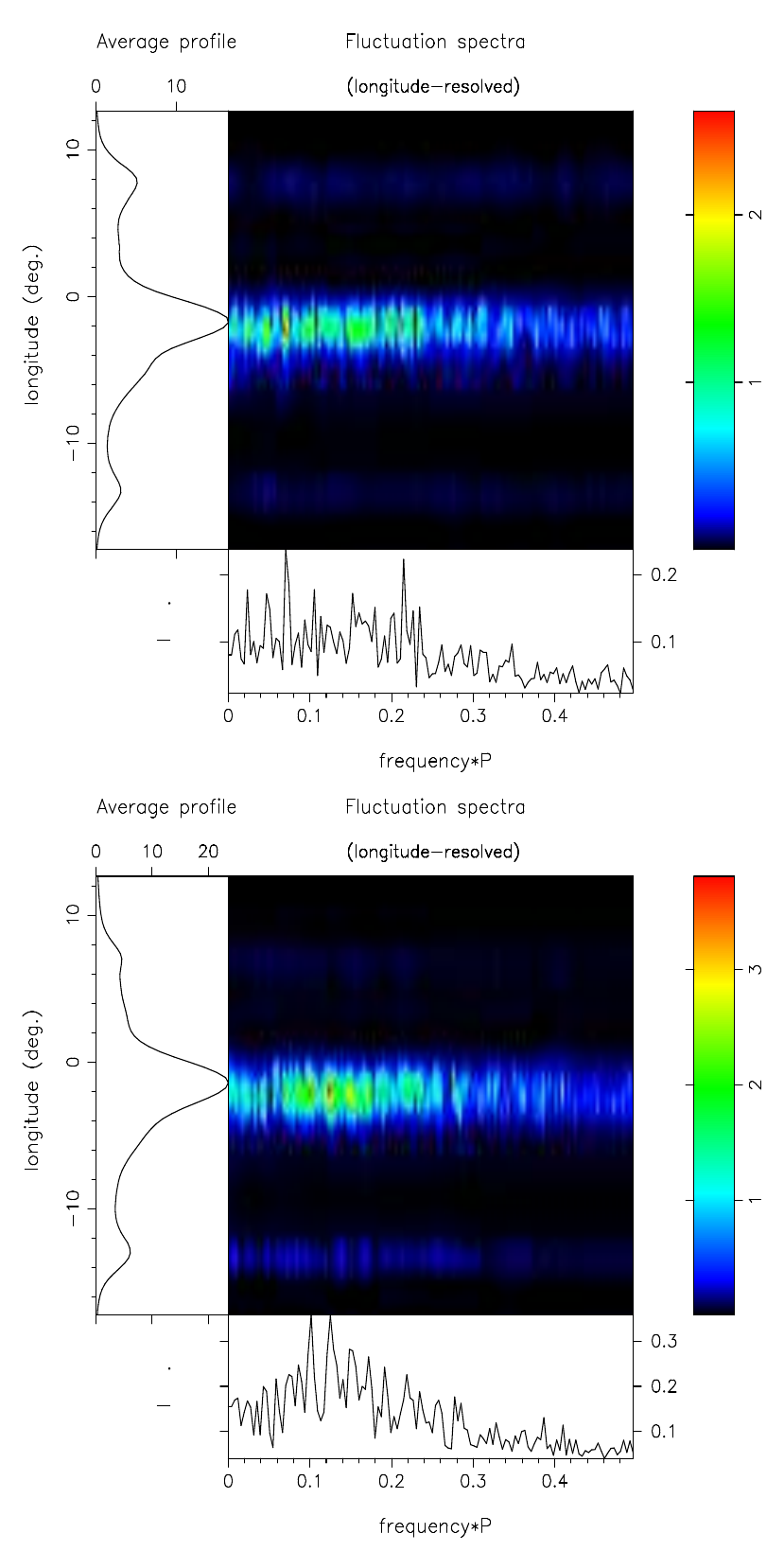}
\caption{Longitude resolved fluctuation spectra (LRFs) of the normal and abnormal modes at 1400 MHz; Normal (upper) {\it vs} abnormal mode (lower).  The center panels give the LRFs, the profile is plotted in the left panels and the integral spectrum at the bottom.}
\end{center}
\label{fig6}
\end{figure}

\begin{table*} 
\begin{tabular}{|l| |l| l |l| l | l | } 
\hline 
\multicolumn{6}{|c|}{Profile Measurements}   \\ 
\hline 
Mode & Profile Component  & $\varphi_{l}$ & $\varphi_{t}$ & $\Delta \varphi$ & r (km)\\
 
\hline 
Normal   & Inner Cone (peak to peak) & $-9.1\pm0.2 \deg$ & $3.4\pm0.2 \deg$ & $-2.6\pm0.2\deg$ & 387$\pm$50 \\
         & Outer Cone (peak to peak) &$-13.4\pm0.2\degr$ & $7.6\pm0.2 \deg$ & $-2.9\pm0.2\deg$ & 431$\pm$50 \\
         & Outer Cone (50\%)         &$-14.9\pm0.3\degr$ & $9.0\pm0.3 \deg$ & $-3.2\pm0.3\deg$ & 476$\pm$70 \\
         & Outer Cone (10\%)         &$-16.5\pm0.3\degr$ &$10.8\pm0.3 \deg$ & $-2.9\pm0.3\deg$ & 431$\pm$70 \\
         & Outer Cone (3$\sigma$)    &$-21.7\pm0.4\degr$ &$18.0\pm0.4 \deg$ & $-1.9\pm0.4\deg$ & 283$\pm$80 \\
         & Core                      &                   &                  & $-1.9\pm0.2\deg$ & 283$\pm$50 \\

Abnormal & Inner Cone (peak to peak) &$-9.2 \pm0.2 \deg$ & $2.9\pm0.2 \deg$ & $-3.2\pm0.2\deg$ & 476$\pm$50 \\
         & Outer Cone (peak to peak) &$-13.6\pm0.2\degr$ & $6.4\pm0.2 \deg$ & $-3.6\pm0.2\deg$ & 535$\pm$50 \\
         & Outer Cone (50\%)         &$-14.7\pm0.3\degr$ & $8.4\pm0.3 \deg$ & $-3.2\pm0.3\deg$ & 476$\pm$70 \\
         & Outer Cone (10\%)         &$-16.2\pm0.3\degr$ &$11.5\pm0.3 \deg$ & $-2.4\pm0.3\deg$ & 357$\pm$70 \\
         & Outer Cone (3$\sigma$)    &$-23.4\pm0.4\degr$ &$20.2\pm0.4 \deg$ & $-1.6\pm0.4\deg$ & 238$\pm$80 \\
         & Core                      &                   &                  & $-1.7\pm0.2\deg$ & 253$\pm$50 \\
\hline
\end{tabular} \\ 
\medskip 
\caption{The profile widths at 10$\%$, 50\%, and 3$\sigma$ intensity are shown for the conal components, along with the angular differences between profile centers and the steepest PPA gradient points ($\Delta \varphi$). For the core component, the change in this angular difference from low to high intensity average profiles provides a $\Delta \varphi$ value with which to calculate the A/R height.}
\label{tab2}
\end{table*} 

The abnormal mode profiles vary much more strongly with frequency than those of the pulsar's normal mode, and we see that the profile forms we have identified as normal and abnormal correspond closely with those identified by Bartel \etal\ (1981).  Table~\ref{tab1} gives the characteristics of the Effelsberg observations for the two modes: their lengths, the average profile widths, and the fractional linear and circular polarizations.  Strikingly, the profile widths for the two modes are very similar.  The average linear polarization in the normal mode is marginally lower than the abnormal mode at 1400 MHz, and the 8.3-GHz profiles are more depolarized.  

The 1.4-GHz abnormal mode observation is longer and of better quality than its 8.3-GHz counterpart.  Also, we can reliably estimate the relative strengths of the ab/normal modes at 1.4 GHz whereas refractive scintillation makes this impossible at 8.3 GHz.  We note the relative strengths of the ab/normal modes at 13 and 3 cm in the five observations of Fig. 3 in Yan \etal\ (2018). The relative strengths of the modes vary substantially from run to run due to their narrow bandwidths, but overall the abnormal mode is seen to be brighter than the normal mode by a factor of the order of 50\%  Therefore, in the remainder of the analysis of the polarization below, we will 
focus on the 1.4-GHz observation for comparison of the two modes because of its length and quality.

\subsection{Polarization Properties of B0329+54 at 1.4 GHz}

We investigated the single pulse polarization properties of both modes. The pulse sequences in each mode were separated into fractions based on the intensity of the core emission using the technique described in MRG07. In Figure 2 we show the pulse polarization plots for the lowest  intensity population (left panels) and the full population (right panels), for each mode. The top panels of each plot represent the average intensity, linear and circular polarization. The bottom panels show the distribution of the single pulse phase-resolved PPA (colour) with only statistically significant points exceeding three times the off-pulse noise level, and the average PPAs are overlaid as a red curve.

We found that overall the full pulse sequences showed little change in their PPA traverses between the two modes, as shown in the right hand displays of Figure 2.  The PPA traverses follow the RVM closely for the low intensity population, whereas at higher intensities the well known non-RVM ``kink'' emerges and the PPA changes more rapidly under the central core component. An overlay of the average polarization properties of the lowest intensity population for 1.4 GHz is shown in the top left panel of Figure 3, where clearly the major difference in linear polarization occurs towards the trailing profile component, whereas the abnormal mode shows a distinct
jump in the PPA where there is a dip in the linear polarization---showing up as a bifurcated feature. The full pulse sequence includes the effect of the core emission, and the PPA traverse below the core component for the normal mode appears to be steeper than the abnormal mode as seen in the overlay plot on the right panel of Figure 3.  The abnormal mode circular peak trails that of the normal mode under the central core component and vanishes in the trailing component.  The average linear and circular polarization appear to change by about 20\% between the two modes (see Table~\ref{tab1}). For comparison, the lower intensity and full profile plots for 8.3 GHz are shown in the bottom panel of 
Figure 3.

Next, using a similar method to MRG07, we fitted the PPA traverse of the lowest intensity pulses to the Rotation Vector Model (RVM), given by
\begin{center}
$\chi=\chi_{\circ}+\tan^{-1}\frac{\sin(\alpha)\sin(\varphi-\varphi_{\circ})}{\sin(\alpha + 
\beta)\cos(\alpha)-\sin(\alpha)\cos(\alpha + \beta)\cos(\varphi-\varphi_{\circ})}$ \\ 
\end{center} 
Here $\alpha$ is the angle between the rotation axis and the magnetic axis, $\beta$ is the angle between the magnetic axis and the observer's line of sight, and $\chi_{\circ}$ and $\varphi_{\circ}$ correspond to the PPA and longitude values of the inflection or the steepest gradient point of the PPA traverse.  We started the fits by fitting the PPA traverse of the normal mode using the $\alpha$ and $\beta$ values obtained by MRG07 as our initial guess, and found that the best fit values
converged to $\alpha=-35.5\degr\pm 2\degr$, $\beta=-2.9\degr \pm 0.5
\degr$, $\chi_{\circ}=-25.3\degr\pm 3\degr$, and we choose
$\varphi_{\circ} = 0\degr\pm 0.3\degr$. The same values were also found to
fit the PPA traverse of the abnormal mode remarkably well.  Therefore,
the basic PPA traverse geometry in the pulsar's abnormal mode appears
identical to that in the normal mode.

It should be noted that the angles $\alpha$ and $\beta$ obtained from
RVM fits are highly correlated (about 98\%) and hence cannot provide
any meaningful constraint to the geometry (von Hoenchbroech \& Xilouris 1997; Everett
\& Weisberg 2001).  However, the quantities
$\chi_{\circ}$, $\varphi_{\circ}$, and the steepest gradient evaluated at
this point $\mid d\chi/d\varphi\mid = \sin(\alpha)/\sin(\beta) \sim11.5$
is significantly better constrained in the fits. 

\subsection{Geometry and Emission Heights}
Since we have determined that the polarization properties are nearly
identical between the two modes, we now proceed to compare the radio emission heights of the two modes using the technique of aberration/retardation (A/R) delay (\eg, Blaskiewicz \etal\ 1991;
Mitra \& Li 2004; Mitra \& Rankin 2011). In this method, if the emission across the profile arises from the same height, and the PPA traverse can be modeled using the RVM, then the negative phase shift
measured between the PPA steepest gradient point and the profile center $\Delta\varphi$ is an emission-height effect, and this allows us to calculate that height using the following relation:
 \begin{equation}
r=-\frac{c\Delta\varphi P}{4*360}
\label{eq1}
\end{equation}
where $P$ is the pulsar rotation period (Dyks \etal\ 2004).  Measurement of $\Delta\varphi$ involves determination of the phase of the leading ($\varphi_l$) and trailing ($\varphi_t$) edges of the emission profile, and for the steepest gradient point defined as zero longitude, $\Delta\varphi =
\varphi_l + (\varphi_t - \varphi_l)/2$. A detailed study for application of
this technique can be found in Mitra \& Li (2004).

PSR B0329+54 has been classified as a multiple profile (Rankin 1993) with inner/outer and central core emission, and MRG07 systematically applied the A/R method to obtain emission heights for all three emission regions. We use the same arguments as MRG07 to find the emission heights for the two modes.  The RVM fits to the lowest intensity pulses allow us to fix the steepest gradient point as the zero longitude.  To find $\varphi_l$ and $\varphi_t$ for the inner/outer cone, it is important to identify the conal emission clearly. 

We found that the peaks of the inner-conal components were more discernible in the lowest intensity profiles, and hence we used this to measure inner-cone peaks as $\varphi_l$ and $\varphi_t$ to find $\Delta\varphi$. For the outer-conal components we used the full profile to find $\varphi_l$ and $\varphi_t$, and subsequently $\Delta\varphi$.  As was pointed out by Mitra \& Li (2004), it is important to use several profile measures to see the effect of A/R; thus we have measured $\varphi_l$ and $\varphi_t$ at 3$\sigma$,
10\%, 50\% and peaks for the outer components of the profile. For the inner cone such an exercise was not possible since only peaks of the components can be identified). The values for $\varphi_l$,
$\varphi_t$, $\Delta\varphi$ and the corresponding emission heights $r$ using
eq.~\ref{eq1} for the inner and outer cones for various profile measures are given in Table~\ref{tab2}. The emission heights measured for the two modes are similar, and the minor differences seen can be
attributed to the uncertainties involved in the assumptions used for applying the A/R method and measurement of $\varphi_l$ and $\varphi_t$ (see Mitra \& Li 2004).

The full intensity profiles were used to determine the location of the peak of the core emission, which was used to find the emission height as given in Table~\ref{tab2}.  Following MRG07, we took the five intensity fractions and plotted them against each other to assess the
shift in their pulse center.  Higher intensity pulses appear progressively earlier due to A/R within the core component if their larger intensity reflects a larger emission height. Therefore, if the core component is A/Red by the same amount in each mode, we have
additional confirmation that the emission height changes little between the modes and thus that the core emission process is unvarying.  The differences in profile center of the different intensity fractions for both modes are shown in Figure 4. The center of the maximum intensity fraction was found to shift by $\sim$ 2$\degr$ in both profile modes, highlighting the lack of change in emission height and the consistent core emission properties.

\subsection{Polarization-mode Segregation}
As discussed in the Introduction, PSR B0329+54's average-profile PPA traverse is quite complex. Although when carefully analyzed at the single pulse level, the PPA traverse can be interpreted as having two orthogonally separated PPA tracks, with each being compatible with the RVM.  These tracks are referred to as the primary and secondary polarization-mode (PPM and SPM)
constituents---the PPM usually meaning the polarization mode dominating the average PPA traverse.  The incoherent addition of the PPM and SPM power then leads to the depolarization observed in the pulsar's average profile.

The PPM and SPM's linearly polarized power distribution is independent of the profile mode changes, and here we provide an additional measure for assessing them by using two methods (see Deshpande \& Rankin 2001).  The first method is called the two-way separation where the corresponding single pulse PPA is first identified with the PPM or SPM, and then the corresponding linear, circular and unpolarized power is associated with a particular polarization-mode.  Here, unpolarized power is assumed to reflect equal PPM/SPM contributions. The second method is the three-way separation where after identification of the PPA with the PPM/SPM, the linear and circular polarized power of the two polarization modes is accumulated, and the unpolarized power is accumulated separately. If the unpolarized power in the three-way separation is significantly small, then it reflects that most of the depolarization in the average pulse profile arises due to incoherent addition of PPM and SPM power.

These two methods of polarization-mode-segregated pulse sequences were computed for the 1.4-GHz observation according to the recipes in Deshpande \& Rankin (2001).  Profiles corresponding to the two-way segregations are shown in Figure 5 where the ab/normal mode profiles are on the bottom/top rows, and the PPM/SPM are on the left/right-hand sides.  The three-way segregations indicated that most of the depolarization was due to incoherent addition of PPM and SPM power, thus they added no new information and are not shown.  We were earlier able to identify the PPM with the X propagation mode and the SPM as the O (Mitra \etal\ 2007), and we use this identification here.

Figure 5 shows that the X/O polarization profiles of the two modes are quite similar.  The principal difference is the truncation of X-mode power on the trailing edge of the profile in the abnormal mode.  This is interesting because the fractions of X-mode power are identical in the normal and abnormal modes at 71\%.  Further, the abnormal mode overall is 1.7 times stronger than the normal mode.

\subsection{Fluctuation Spectra}
Given that the modes of some pulsars exhibit periodic fluctuations, it seemed important to assess whether this might be so for B0329+54's abnormal mode.  Longitude resolved fluctuation spectra (LRFs) are
given in Figure 6 for the pulsar's normal (upper) and abnormal (lower) modes at 1400 MHz.

A number of LRFs have been published (e.g. Taylor \& Huguenin 1971, Weltevrede \etal\ 2006/2007, Yan \etal\ 2018) that do not distinguish between the pulsar's modes---however, these will be dominated by the normal mode and show a broad low frequency feature peaking at about 8 c/$P$.
Unfortunately, our normal mode LRFs seem to be corrupted by RFI, but they do show fluctuation power in this interval.  The abnormal mode LRFs, however, show a character of fluctuation power identical to that seen in published LRFs (Yan \etal\ 2018); therefore we can conclude that there is little or no difference between the modes in terms of modulation periodicity.

\section{Results and Discussion}
  
We have determined that the normal and abnormal modes of pulsar B0329+54 exhibit the same basic emission geometry in terms of $\alpha$ and $\beta$ from RVM solutions as well as emission altitudes measured using the A/R method (given in Table 2). Even the same PPA "kink" is seen in the high intensity pulses of each profile mode. Only moderate changes in emission properties suggest that nothing about the overall emission geometry or dipolar magnetic configuration changes from the abnormal mode to the normal: therefore these cannot be responsible for the mode switching. This conclusion is consistent with the findings of Bartel \etal\ (1982)
 where they also report moderate changes in the pulse shape and polarization properties during mode changes.

It is generally accepted that the coherent radio emission from pulsars arises due to growth of plasma instabilities in relativistically streaming electron-positron plasma along the dipolar open magnetic field lines. At a height of around a few hundred kilometers, where the radio emission detaches from the magnetosphere, the magnetic field is significantly strong and the only plasma instability that can be generated is the two-stream instability. The coherent radio emission model of Ruderman \& Sutherland (1975) suggests that an inner accelerating region can develop above the polar cap where electron-positron pairs can be created, eventually leading to a non-stationary electron-positron plasma flow in the form of several localized sparks. In this model, the growth of two-stream instability and its non-linear evolution can lead to the formation of charged bunches (relativistic solitons) which can excite coherent curvature radiation along the curved magnetic fields (Mitra \& Basu 2018; Asseo \& Melikidze 1998; Melikidze et al. 2000; Gil \etal\ 2004; Mitra, Gil \& Melikidze 2009, Lakoba, Mitra \& Melikidze 2018) at a characteristic frequency of 
\begin{center}
$\nu_c \sim \gamma^3 c/\rho_c$
\end{center}
where $\gamma$ is the Lorentz factor of the charged soliton, $\rho_c$ the radius of curvature of the magnetic field and $c$ is the velocity of light.  The power radiated from the charged soliton is 
\begin{center}
$P \sim F_Q \gamma^4/{\rho_c}^2$
\end{center} 
where $F_Q$ is a function of plasma parameters and has the dimension of charge squared (see Melikidze et al. 2000; Basu and Mitra 2017).  The emission from a large number of such charged bunches adds up incoherently to give rise to the observed radio intensity (Asseo \& Melikidze 1998; Melikidze et al. 2000; Gil et al. 2004). In this model, the spark-associated plasma column corresponds to the observed subpulses in single pulses and sometimes components in a pulse profile.

Let us now apply the soliton curvature radiation model to the normal and abnormal modes of PSR B0329+54.  Here, we have established that the overall emission at 1.4 GHz across the profile arises from similar heights for both the normal and abnormal modes, and hence the underlying $\rho_c(\varphi)$ is same for both the modes, where the radius of curvature $\rho_c$ depends on the longitude $\varphi$. This in turn suggests that to obtain same $\nu_c(\varphi)$ for the two modes, $\gamma_c(\varphi)$ should also be similar, and hence the ratio of the power emitted by the two modes should be 
\begin{center}
$\frac{P(\varphi)^{abnormal}}{P(\varphi)^{normal}} \sim \frac{F_Q(\varphi)^{abnormal}}{F_Q(\varphi)^{normal}}$
\end{center}
The top panel in Fig. 1 shows that on average the power in the abnormal mode is roughly 1.7 times more than in the normal mode, and hence we can infer that for longitude averaged pulses: 
\begin{center}
$F_Q^{abnormal} \sim 1.7F_Q^{normal}$
\end{center}
Assuming as above that the modes differ in terms of the amount of charge or pair plasma involved in the radiative processes, it then seems to follow that power in the X/O modes is proportional as the power increases, but not uniform in its effects within the pulsar's core and conal beams.  At 1400 MHz the trailing X-mode power
decreases, but at higher frequencies it increases as is shown in Fig. 1, so the change is in the trailing component spectrum, not overall power. Despite this spectral difference, the overall abnormal mode power is roughly 70$\%$ stronger than the normal mode, and the X and O modal powers increase proportionately as shown in
Figure 5.

The charge available to produce emission comes from electron/positron pair-production cascades. Changes in this production or cascade of pair plasma can result in differences in profile shape without altering the emission height, or the open magnetic field line geometry. In inner vacuum gap models (Ruderman \& Sutherland 1975), as discussed earlier, a potential gap forms on the magnetic polar cap
that gets discharged by pair production. As this discharge proceeds around the magnetic axis, it changes the electrostatic conditions in the sparking zone. 

If a source could provide additional photons at the very beginning of this process in a quasi-periodic fashion, these
metastable plasma modes could produce regular and stable differences in profile shape. This additional source could, in principal, arise from a region outside the polar gap; alternatively, changes to the magnetic field within the polar cap could account for a differing quantity of charge along the electron cascade process. Quasi-periodic fluctuations of these closed field lines---without changing the open magnetic field region from where the radiation detaches---could lead to distinct metastable states of plasma flow and hence explain profile mode changes.

\section{Conclusion}
We investigated the polarization characteristics of one pulsar, B0329+54, in 
order to isolate the most probable mechanism for mode switching.  We found 
that the only possible change that could produce a difference in the power of 
the normal and abnormal modes is a change in the group charge Q.  Indeed, we found that the overall intensity of the abnormal mode is almost twice that of the normal mode; whereas the radiated X/O power ratio is nearly identical between the two modes. Thus the trailing component attenuation at 1.4 GHz (and increase at 8.3 GHz) is a spectral rather than a power effect. 

Further modeling is needed to show what produces this change along the electron cascade process. There could be several possible explanations for what could change the surface energy, therefore providing a different number of photons for pair production; but a wider survey of more profile mode-switching pulsars could help illuminate what exactly causes the pair production to change.

\section*{Acknowledgments} We thank Ramesh Karuppusamy for providing us the software necessary to read the archival data formats. We also thank Michael Kramer for providing access to the Effelsberg archival observations used in this work. Portions of this work were supported by US NSF grants 09-68296 and 18-14397.  This work made use of the NASA ADS astronomical data system. DM acknowledges funding from the grant
``Indo-French Centre for the Promotion of Advanced Research - CEFIPRA".

\bibliographystyle{mnras}

\begin{thebibliography}{mnras}
\bibitem[\protect\citeauthoryear{Asseo \& Melikidze}{1998}]{a1} Asseo, E., \& Melikidze, G.I. 1998, \mnras, 301, 59A
\bibitem[\protect\citeauthoryear{Backer}{1970}]{n1} Backer, D. C. 1970, \nat, 228, 42
\bibitem[\protect\citeauthoryear{Backer \& Rankin}{1980}]{n2} Backer D.C., \& Rankin, J.M. 1980, \apjs, 42, 143
\bibitem[\protect\citeauthoryear{Bartel}{1981}]{n64} Bartel N. 1981, \aap, 97, 384
\bibitem[\protect\citeauthoryear{Bartel \etal\ }{1982}]{n3} Bartel N., Morris D., Sieber W., \& Hankins, T.H. 1982, \apj, 258, 776
\bibitem[\protect\citeauthoryear{Basu \& Mitra}{2018}]{bm18} Basu, R., \& Mitra, D. 2018, \mnras, 476, 1345
\bibitem[\protect\citeauthoryear{Bia\l{}kowski \etal\ }{2018}]{n7} Bia\l{}kowski, S., Lewandowski, W., Kijak, J., 
	B\l{}aszkiewicz, L., Krankowski, A., Os\l{}owski, S. 2018, Ap\&SS, 363, 110

\bibitem[\protect\citeauthoryear{Blaskiewicz, Cordes \& Wassermann}{1991}]{n5} Blaskiewcz, M., Cordes, J.M., \& Wassermann, I. 1991 \apj, 370 643.
\bibitem[\protect\citeauthoryear{Chen \etal\ }{2011}]{c3} Chen, J. L., Wang, H. G., Wang, N., Lyne, A., Liu, Z. Y., Jessner, A., Yuan, J. P., \& Kramer, M. 2011, \apj, 741, 48

\bibitem[\protect\citeauthoryear{Cordes}{2013}]{co3} Cordes, J. M. 2013, \apj, 775, 47
\bibitem[Deshpande \& Rankin (2001)]{dr01} Deshpande, A.A. \& Rankin, J.M. 2001, \mnras, 322, 438
\bibitem[\protect\citeauthoryear{Dyks \etal\ }{2004}]{n8} Dyks, J., Rudak, B., \& Harding, A.K. 2004, \apj, 607, 939
\bibitem[\protect\citeauthoryear{Edwards \& Stappers}{2004}]{n9}  Edwards, R.T., \& Stappers, B.W. 2004, \aap, 421, 681 (ES04).  
\bibitem[\protect\citeauthoryear{Everett \& Weisberg}{2001}]{n10}  Everett, J. E., \& Weisberg, J. M. 2001, \apj, 553, 341.
\bibitem[\protect\citeauthoryear{Gangadhara}{1995}]{n11}  Gangadhara, R. T. 1995, \apss, 232, 327.
\bibitem[\protect\citeauthoryear{Gangadhara \& Gupta}{2001}]{n12} Gangadhara R.T., Gupta Y.2001, \apj, 555, 31(G\&G)
\bibitem[\protect\citeauthoryear{Gil \& Lyne}{1995}]{n14}  Gil, J. A., \& Lyne, A. G. 1995, \mnras, 276, 55.
\bibitem[\protect\citeauthoryear{Gil \etal\ }{2004}]{g04}  Gil, J. A., Lyubarsky, Y., \& Melikidze, G.I. 2004, \apj, 600, 872G.
\bibitem[\protect\citeauthoryear{Gould \& Lyne}{1998}]{n15}  Gould, D.M., \& Lyne, A.G. 1998, \mnras, 301, 253.
\bibitem[\protect\citeauthoryear{Gupta \etal\ }{2000}]{g16}  Gupta, Y.; Gothoskar, P., Joshi, B. C., Vivekanand, M.; Swain, R., Sirothia, S.\& Bhat, N. D. R., 2000, ASP Conf. Ser. 202: 
\bibitem[\protect\citeauthoryear{von Hoensbroech \& Xilouris}{1997}]{n19}  von Hoensbroech, A., \& Xilouris, K.M. 1997, \aaps, 126, 121.
\bibitem[\protect\citeauthoryear{von Hoensbroech}{1999}]{n20}  von Hoensbroech, A. 1999, Ph.D. Thesis, Bonn University, Bonn, Germany.
\bibitem[\protect\citeauthoryear{Jessner}{1996}]{j96}  Jessner, A. 1996, Large Antennas in Radio Astronomy proceedings p.185.

\bibitem[\protect\citeauthoryear{Johnston \etal\ }{2005}]{n48} Johnston, S., Hobbs, G., Vigeland, S., Kramer, M., Weisberg, J. M., \& Lyne, A. G.  2005, \mnras, 364, 1397
\bibitem[\protect\citeauthoryear{Kijak \& Gil }{1997}]{kg97}{Kijak} J., {Gil} J., 1997, \mnras, 288, 631
\bibitem[\protect\citeauthoryear{Komesaroff}{1970}]{n21} Komesaroff, M.M. 1970, \nat, 225, 612
\bibitem[\protect\citeauthoryear{Kramer \etal\ }{1999}]{n16}  Kramer, M., Xilouris, K. M., Camilo, F., Nice, D. J., Backer, D. C., Lange, C., Lorimer, D. R., Doroshenko, O., \& Sallmen, S. 1999, \apj, 520, 324
\bibitem[\protect\citeauthoryear{Kramer \etal\ }{2006}]{p16} Kramer, M., Lyne, A. G., O'Brien, J. T., Jordan, C. A., \& Lorimer, D. R. 2006, \sci, 312, 549.
\bibitem[\protect\citeauthoryear{lakoba}{2018}]{l21} Lakoba, T., Mitra, D. \& Melikidze, G., 2018,  \mnras, 480, 4526
\bibitem[\protect\citeauthoryear{Lyne}{1971}]{ly21} Lyne, A.G. 1971, \mnras, 153, 27
\bibitem[\protect\citeauthoryear{Mahajan \etal\ }{2018}]{m21} Mahajan, N., van Kerkwijk, M. H., Main, R., \& Pen, U.-L. 2018, arXiv180701713M

\bibitem[\protect\citeauthoryear{Malov \& Suleymanova}{1998}]{n23} Malov, I. F., \& Suleymanova, S.A. 	1998, Astron. Rep. 42, 388 (MS98).

\bibitem[\protect\citeauthoryear{Melikidze \etal\}}{2000}]{M00} Melikidze, G.I., Gill, J.A., \& Pataraya, A.D., 2000, \apj, 544, 1801M. 

\bibitem[\protect\citeauthoryear{McKinnon \& Hankins}{1993}]{n27}  McKinnon, M. M., \& Hankins, T. H. 1993,  \aap, 269, 325 (MH93).
\bibitem[\protect\citeauthoryear{Mitra \& Li}{2004}]{m27}  Mitra, D., \& Li, X. H. 2004, \aap, 421, 215
\bibitem[\protect\citeauthoryear{Mitra \etal\ }{2007}]{m49} Mitra, D., Rankin, J. M. \& Gupta, Y. 2007, \mnras, 379, 932 (MRG07)
\bibitem[Mitra \etal\ (2009)]{mit09} Mitra, D.; Gil, J.; Melikidze, G.  2009, ApJL, 696, L141
\bibitem[\protect\citeauthoryear{Mitra \& Rankin}{2011}]{r11}  Mitra, D., \& Rankin, J. M. 2011, \apj, 727, 92M 
\bibitem[Mitra \etal\ (2016)]{mit16} Mitra, D.; Basu, R.; Maciesiak, K.; Skrzypczak, A.; Melikidze, G.I.; Szary, A.; Krzeszowski, K.  2016, ApJ, 833, 28.
\bibitem[\protect\citeauthoryear{Morris \etal\ }{1979}]{n49} Morris, D., Graham, D. A., Sieber, W., Jones, B. B., Seiradakis, J. H.,  \& Thomasson, P. 1979, \aap, 73, 46
\bibitem[\protect\citeauthoryear{Radhakrishnan \& Cooke}{1969}]{n33} Radhakrishnan, V., \& Cooke, D. J. 1969, Ap. Lett, 3, 225
\bibitem[\protect\citeauthoryear{Rankin}{1983}]{n36} Rankin, J.M. 1983a, \apj, 274 333
\bibitem[\protect\citeauthoryear{Rankin}{1983}]{n37} Rankin, J.M. 1983b, \apj, 274 359
\bibitem[\protect\citeauthoryear{Rankin}{1986}]{n38} Rankin, J.M. 1986, \apj, 301, 901
\bibitem[\protect\citeauthoryear{Rankin}{1990}]{n39} Rankin, J.M. 1990, \apj, 352, 247 (R90)
\bibitem[\protect\citeauthoryear{Rankin}{1993}]{n40} Rankin, J.M. 1993, \apj, 405, 285/\apjs, 85, 145
\bibitem[\protect\citeauthoryear{Rankin}{2007}]{r07} Rankin, J.M. 2007, \apj, 664, 443.
\bibitem[\protect\citeauthoryear{Rankin}{2015}]{r15} Rankin, J.M. 2017, \apj, 804, 112.
\bibitem[\protect\citeauthoryear{Ruderman \& Sutherland}{1975}]{r75}  Ruderman, M.A., \& Sutherland, P.G. 1975, \apj, 727, 92M 
\bibitem[\protect\citeauthoryear{Smits \etal\ }{2005}]{n41} Smits, J. M., Mitra, D., \& 
	Kuijpers, J. 2005, \aap, 196, 51R
\bibitem[\protect\citeauthoryear{Taylor \& Huguenin (1971)}{2017}]{th71}Taylor, J. H.\& Huguenin, G. R., 1971, \apj, 167,273.

\bibitem[\protect\citeauthoryear{Timokhin}{2010}]{t10} Timokhin, A. N. 2010 \mnras, 408L, 41
\bibitem[\protect\citeauthoryear{Yan \etal\ }{2018}]{ym18}Yan, Z., Shen, Z.-Q., Manchester, R. N., Ng, C.-Y., Weltevrede, P., Wang, H.-G., Wu, X.-J., Yuan, J.-P., Wu, Y.-J., Zhao, R.-B., Liu, Q.-H., Zhao, R.-S., Liu, J. 2018, \apj, 856, 55

\bibitem[\protect\citeauthoryear{Yuen \& Melrose}{2017}]{ym17} Yuen, R., \& Melrose, D. B. 2017, \mnras, 469, 2049

\bibitem[Weltevrede \etal\ (2006)]{wel06} Weltevrede, P., Edwards, R. T., \& Stappers, B. W.  2006, \aap, 445, 243
\bibitem[Weltevrede \etal\ (2007)]{wel07} Weltevrede, P., Stappers, B. W., \& Edwards, R. T.  2007, \aap, 469, 607

\end{thebibliography}

\label{lastpage}

\end{document}